\documentclass[twocolumn]{aastex63}

\def\lapp{\ifmmode\stackrel{<}{_{\sim}}\else$\stackrel{<}{_{\sim}}$\fi}
\def\gapp{\ifmmode\stackrel{>}{_{\sim}}\else$\stackrel{>}{_{\sim}}$\fi}
\usepackage{multirow}
\usepackage{color}
\usepackage{amsmath}
\usepackage{soul}
\usepackage{hyperref}
\usepackage{lineno}
\shorttitle{}
\shortauthors{}
\begin{document}
\title{Orbital Modulation of Gamma Rays from PSR~J2339$-$0533}

\correspondingauthor{Hongjun An}
\email{hjan@cbnu.ac.kr}

\author{Hongjun An}
\affiliation{Department of Astronomy and Space Science,
Chungbuk National University, Cheongju, 28644, Republic of Korea}
\author{Roger W. Romani}
\affiliation{Department of Physics/KIPAC, 
Stanford University, Stanford, CA 94305-4060, USA}
\author{Matthew Kerr}
\affiliation{Space Science Division,  
Naval Research Laboratory, Washington, DC 20375-5352, USA}
\collaboration{3}{(Fermi-LAT collaboration)}

\begin{abstract}
We report on orbital modulation of the 100--600\,MeV gamma-ray emission of the $P_{\rm B}=4.6$\,hr millisecond pulsar binary PSR~J2339$-$0533 using 11\,yr of {\it Fermi} Large Area Telescope data. The modulation has high significance (chance probability $p\approx 10^{-7}$), is approximately sinusoidal, peaks near pulsar superior conjunction, and is detected only in the low-energy 100--600\,MeV band.  The modulation is confined to the on-pulse interval, suggesting that the variation is in the  2.9-ms {\it pulsed} signal itself.  This contrasts with the few other known systems exhibiting GeV orbital modulations, as these are unpulsed and generally associated with beamed emission from an intrabinary shock.  The origin of the modulated pulsed signal is not  yet clear, although we describe several scenarios, including Compton upscattering of photons from the heated companion. This would require high coherence in the striped pulsar wind.
\end{abstract}

\keywords{binaries: close --- gamma rays: stars --- X-rays: binaries
--- stars: individual (PSR~J2339$-$0533)}

\section{Introduction} \label{sec:intro}

Since the launch of the {\it Fermi} Large Area Telescope \citep[LAT;][]{fermimission}, the list of millisecond pulsar-binary systems is growing rapidly thanks to its wide field of view, large effective area, and continuous all-sky monitoring.  The so-called `spider' binaries, black widows (BWs; with a $<0.1 M_\odot$ companion) and redbacks (RBs; with a $0.1-0.7 M_\odot$ companion),  in which the pulsar wind is evaporating the companion, are believed to be descendants of low-mass X-ray binaries whose neutron star primaries have been spun up over long times by accretion \citep[][]{acrs82}. 

In these systems, the spindown flux of the pulsar heats the pulsar-facing side of the companion, which manifests as day-night cycles in the optical light curves, and a wind from the companion's surface collides with the relativistic pulsar wind to form an intrabinary shock (IBS).
Pair particles energized in the shocked pulsar wind, likely via shock-driven reconnection, accelerate along the contact discontinuity to mildly relativistic velocities and beam synchrotron radiation in a hollow cone pattern \citep[e.g.,][]{rs16,whvb+17,kra19}. When the pulsar momentum flux dominates (as is typical for black widows), the IBS wraps around the companion, so that beamed IBS emission reaches the observer at pulsar superior conjunction (companion in front, optical `night side', binary phase $\phi_{\rm B}=0.25$ with respect to the pulsar ascending node). For the redback case, the wind from the more massive companion generally dominates, the shock wraps around the pulsar, and IBS emission should be centered on pulsar inferior conjunction ($\phi_{\rm B}=0.75$). This synchrotron emission is quite bright in the X-ray band, giving rise to a characteristic double-peak orbital modulation in the X-ray light curve \citep[e.g.,][]{hkth+12} as the observer's line of sight cuts through this hollow cone with the binary rotation. The X-ray modulation is sensitive to the system geometry (e.g., wind strengths, inclination) and thus can supplement the optical modeling \citep[e.g.,][]{kra19} in inferring binary properties.

If sufficiently energetic, IBS particles can also produce gamma rays via synchrotron or inverse Compton emission. However, the very bright GeV pulsar magnetospheric emission dominates, making this signal difficult to detect, and only a handful of such detections have been made \citep[][]{wtch+12,arjk+17,ark18,ntsl+18}. 
Hard spectrum $\gamma$-ray modulation is likely due to inverse Compton scattering, while in the case of PSR J2241$-$5236 the emission is quite soft, with a sub-GeV cutoff, suggesting we may be seeing the upper limit of a beamed synchrotron component \citep{ark18}.

PSR~J2339$-$0533 (J2339 hereafter) is a 2.9-ms pulsar in a 0.19-day orbit with a 0.3$M_\odot$ companion \citep[][]{rs11,khc+12}. It was identified as a pulsar binary by targeted optical studies of the brightest unidentified {\it Fermi}-LAT sources with pulsar-like emission, and then confirmed via LAT and radio pulsations \citep[][]{pppb+20,pc15}.
Its optical light curve shows strong heating, and pulsar timing characterizes it as a redback with a companion mass $M_2 \approx 0.3M_\odot$.  In the X-ray band, the spectrum is a hard power law, and the light curve shows double-peaked structure bracketing $\phi_{\rm B}=0.75$; these are good signatures of RB-class IBS emission \citep[][]{rs11, kra19}. With a GeV flux $\approx 3 \times 10^{-11}\, {\rm erg\,cm^{-2}\,s^{-1}}$, this bright source invites a deep search for gamma-ray modulation, using a long 11\,yr set of {\it Fermi}-LAT data. We show data analyses and the results in Section~\ref{sec:sec2}. We then present toy models in an attempt to describe the modulation in Section~\ref{sec:sec3}; these are not yet fully satisfactory. We conclude in Section~\ref{sec:sec4}. Uncertainties are at the 1$\sigma$ confidence level unless noted otherwise.

\section{Observational Data and Analyses} \label{sec:sec2}

\subsection{Data reduction} \label{sec:sec2_1}

{\it Fermi}-LAT data collected between 2008 Aug 04 and 2019 Jul 28 are downloaded from the public archive\footnote{https://fermi.gsfc.nasa.gov/ssc} and analyzed with the {\it Fermi} Science Tools (ST) version 1.0.1 along with the most recent instrument response files ({\tt P8R3\_SOURCE\_V2}). We select the {\tt Front+Back} event type in the {\tt SOURCE} class in an $R=10^\circ$ region of interest (RoI) centered at the source position. The data are further reduced by requiring the zenith angle $<90^\circ$, DATA\_QUAL$>$0, and LAT\_CONFIG=1. Note that intense solar flares occurred when the Sun was $\sim$7$^\circ$ away from J2339, and we have removed the flare time periods (Section~\ref{sec:sec2_2}).

\begin{figure}
\centering
\includegraphics[width=3.2 in]{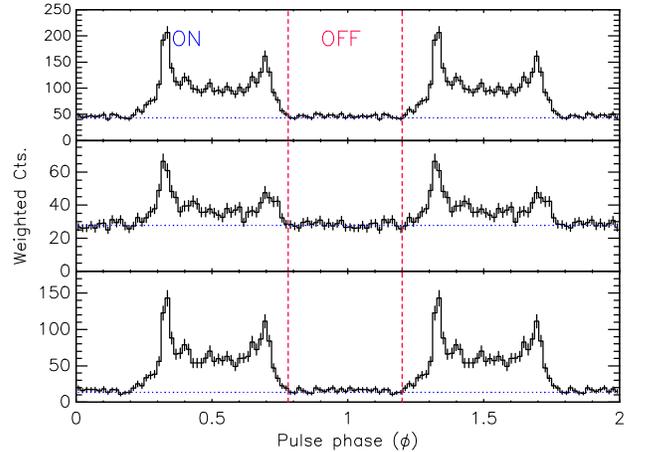} \\
\figcaption{{\it Fermi}-LAT gamma-ray pulse profiles of J2339 from an $R=3^\circ$ RoI in three
energy bands: 0.1--500\,GeV (top), 0.1--0.6\,GeV (middle), and 0.6--500\,GeV (bottom).
Red vertical lines show the off-pulse ($\phi=0.78-1.2$) interval used in our analyses and blue lines denote the average background level estimated using the photon weights: $\sum_i(w_i - w_i^2)/N_{\rm bin}$.
\label{fig:fig1}
}
\vspace{0mm}
\end{figure}

\subsection{Timing Analysis}
\label{sec:sec2_2}

\begin{table}
\begin{center}
\caption{Timing Parameters for PSR~J2339$-$0533}
\label{ta:ta1}
\begin{tabular}{ c c }
\hline
\hline
\rule{0pt}{3ex}
RA ($\alpha$, J2000)              & $23^\mathrm{h}39^\mathrm{m}38\fs741$\\
\rule{0pt}{3ex}
DEC ($\delta$, J2000)            & $-5^\circ  33' 05\farcs108$\\
\rule{0pt}{3ex}
Epoch (MJD)             & 55792 \\
\hline
\rule{0pt}{3ex}
$\nu$ (s$^{-1}$) & 346.71337922047(2) \\
\rule{0pt}{3ex}
$\dot \nu$ (s$^{-2}$) & $-$1.6945(2)$\times 10^{-15}$ \\
\rule{0pt}{3ex}
TZRMJD & 56552.104208432663877 \\
\rule{0pt}{3ex}
Binary model  & ELL1  \\
$F_{\rm B}$ (s$^{-1}$)    & 5.99387361(1)$\times 10^{-5}$ \\
$F_{\rm B1}$ (s$^{-2}$)   & 6.93(6)$\times 10^{-19}$ \\
$F_{\rm B2}$ (s$^{-3}$)   & 1.33(3)$\times 10^{-26}$ \\
$F_{\rm B3}$ (s$^{-4}$)   & $-1.72(2)\times 10^{-33}$ \\
$F_{\rm B4}$ (s$^{-5}$)   & $-2.4(1)\times 10^{-41}$ \\
$F_{\rm B5}$ (s$^{-6}$)   & 5.00(8)$\times 10^{-48}$ \\
$F_{\rm B6}$ (s$^{-7}$)   & $-5.6(4)\times 10^{-56}$ \\
$F_{\rm B7}$ (s$^{-8}$)   & $-1.02(2)\times 10^{-62}$ \\
$F_{\rm B8}$ (s$^{-9}$)   & 5.4(1)$\times 10^{-70}$ \\
$F_{\rm B9}$ (s$^{-10}$)  & $-1.07(5)\times 10^{-77}$ \\
$F_{\rm B10}$ (s$^{-11}$) & 6.8(8)$\times 10^{-86}$ \\
$A1$  ($lt$-$s$) & 0.611668(3) \\
ESP1 & 0 \\
ESP2 & 0 \\
$T_{\rm ASC}$ & 55791.9182100(4) \\
PMRA  ($\dot{\alpha}\cos\delta$, mas\, $yr^{-1}$) & 4.147 \\
PMDEC ($\dot{\delta}$, mas\, $yr^{-1}$) & $-10.311$ \\
\hline
\end{tabular}\\
\end{center}
\footnotesize{{\bf Notes.} 1-$\sigma$ uncertainties are shown in brackets, and parameters without the uncertainty are held fixed. $F_{\rm Bx}$'s are $x$-th time derivatives of the orbital frequency $F_{\rm B}$.\\}
\end{table}

\begin{figure*}
\centering
\begin{tabular}{cc}
\includegraphics[width=3.2 in]{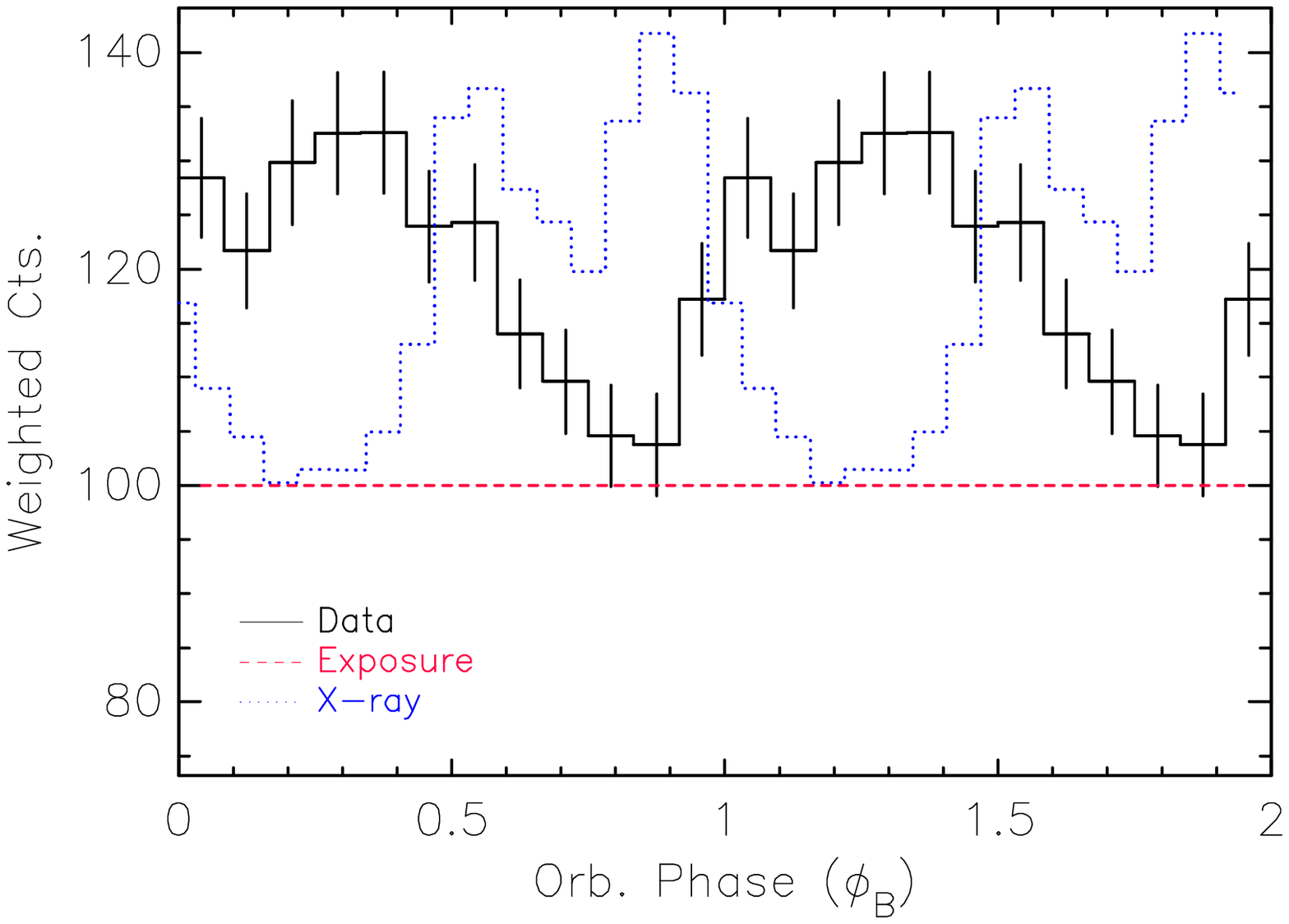} &
\includegraphics[width=3.2 in]{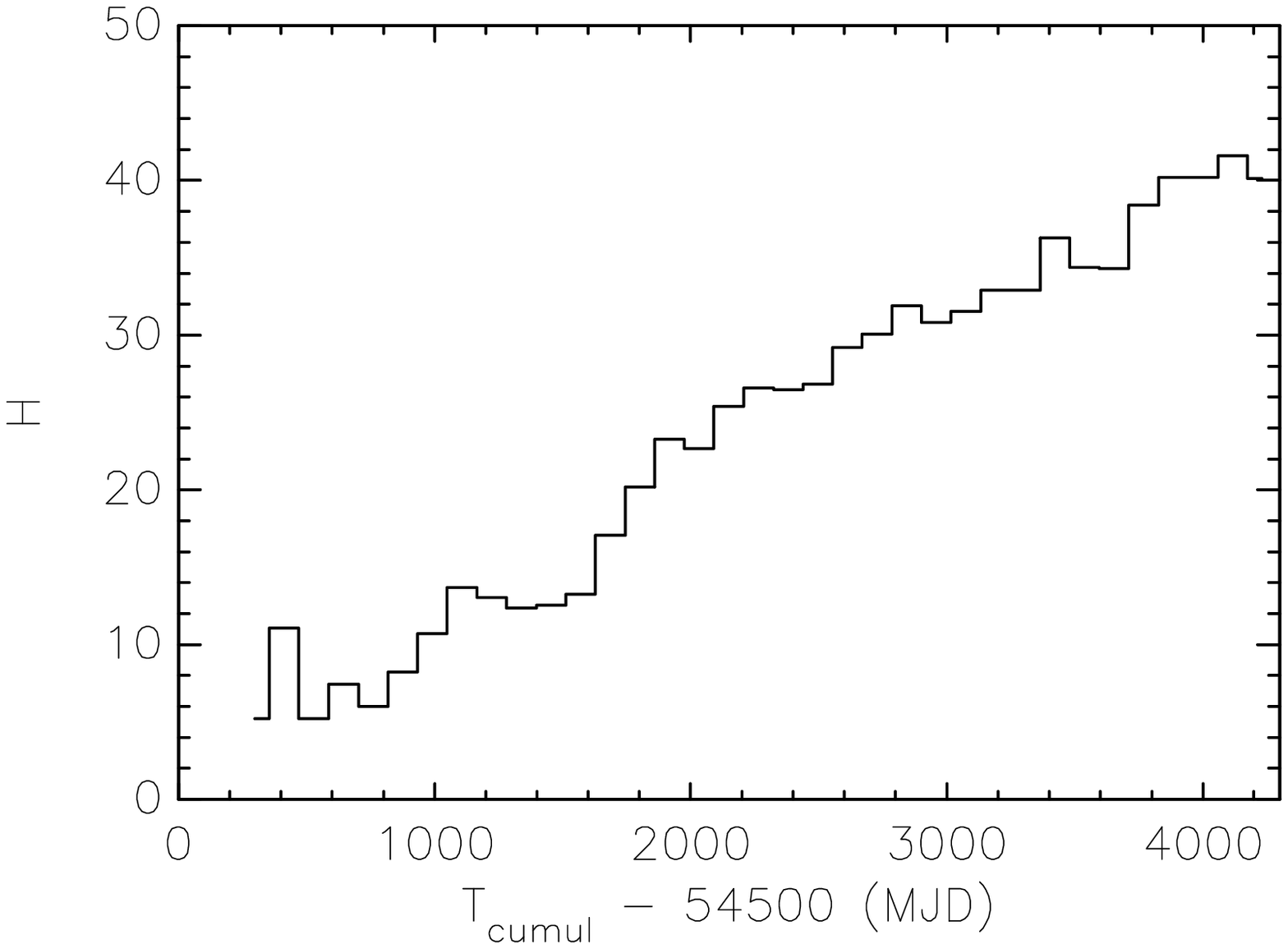} \\
\end{tabular}
\figcaption{Left: 100--600\,MeV orbital light curve in the on-pulse ($\phi=0.20-0.78$) interval. The 3--15\,keV X-ray light curve ({\it Chandra}, {\it XMM-Newton}, and {\it NuSTAR} combined), associated with IBS emission, is shown in blue for reference \citep[][]{kra19}. The X-ray light curve is normalized arbitrarily, and the error bars ($\sim$10\% level) are not shown for better legibility. The red horizontal line in the left panel shows the folded exposure, normalized to a mean value of 100 and with the same binning. Right: the accumulation of $H$-test significance over time.
\label{fig:fig2}
}
\vspace{0mm}
\end{figure*}

J2339 shows significant variability in its binary period, which complicates  long-term analysis of the pulsed signal. Such variability is common in redbacks, but is particularly  strong for J2339. \citet{pc15} analyzed $\sim 6$yr of LAT data to describe the variability and proposed that it is associated with a variable companion quadrupole moment. Since the variability is stochastic, their model does not extend to later times; thus our analysis first requires an updated pulsar timing solution.  For each event, we compute a probability weight $w_i$ \citep[][]{k11} using the {\tt gtsrcprob} tool of {\it Fermi}-ST based on the 4FGL model \citep[][]{fermi4fgl}, and we fold the events using {\tt tempo2} \citep[][]{ehm06} on an initial timing solution to calculate the arrival phases ($\phi$). We then gradually increase the time coverage to 11\,years and generate a new timing solution by maximizing the unbinned likelihood
\begin{equation}
\label{eq:loglike}
\log\mathcal{L}=\sum_i \log [w_i f(\phi(\lambda, t_i)) + 1-w_i],
\end{equation}
where $f$ is an analytic pulse-profile model and $\lambda$ is the set of shape parameters. We use the {\tt PINT} software package\footnote{{https://github.com/nanograv/PINT}} \citep[][]{lrdr+18}, and adjust the timing parameters, holding the eccentricity fixed at 0. For the position and proper motions we use the {\it GAIA} results \citep[][]{gaia18,jkcc+18}. The best-fit timing solution and the parameter uncertainties (1-$\sigma$) are reported in Table~\ref{ta:ta1}, and the resulting pulse profiles are presented in Figure~\ref{fig:fig1}.

Next we investigate orbital modulation of the source. We inspect the $R=3^\circ$ source light curve in three energy bands (100--1000\,MeV, 100--600\,MeV, and 1000--100000\,MeV) using the $H$ test \citep[][]{k11} and find strong modulation in the pulse-phase-summed low-energy (100--600\,MeV) data with $H=27$ corresponding to $p\approx2\times10^{-5}$; higher-energy modulation is insignificant with $H\approx 1$. Note that the two low-energy bands are not independent and accounting for two trials would not affect these results significantly. We find that this low-energy modulation peaks near $\phi_{\rm B} \approx 0.25$. This is surprising, as for typical RB parameters we expect an IBS maximum to occur near $\phi_{\rm B}=0.75$; indeed during the X-ray observations the IBS clearly brackets this phase, implying that (at least at these times) the wind wraps around the pulsar.

As noted above, in the gamma-ray band IBS emission is far fainter than the strong pulsed magnetospheric emission. Thus we would expect that IBS orbital modulation would be stronger in spin phases where the magnetospheric emission is beamed away from Earth. We thus define on- ($\phi=0.2-0.78$) and off-pulse ($\phi=0.78-1.2$)  intervals (Fig.~\ref{fig:fig1}) and generate pulse-phase-selected orbital light curves in the 100--600\,MeV band using the $R=3^\circ$ RoI. Surprisingly, we find that modulation in the ``on-pulse'' interval is very strong with $H\approx 40$ ($p\approx 10^{-7}$; Fig.~\ref{fig:fig2} left) while that in the off-pulse interval is insignificant with $H\approx 1$. Changing the RoI size and/or pulse-phase selection slightly does not alter the results, but they are relatively sensitive to the energy selection; increasing the higher-energy bound reduces the significance (e.g., $p\approx 2\times10^{-4}$ in the 100\,MeV--1\,GeV band). We further verified that the significance of the 100--600\,MeV modulation increases approximately monotonically with time (Fig.~\ref{fig:fig2} right).

We test whether this modulation is induced by exposure variation or flares of nearby sources (e.g., blazars). We compute 30-s binned exposure  using the {\tt gtexposure} tool of {\it Fermi}-ST and fold the exposure using the same timing solution (Table~\ref{ta:ta1}); the exposure variation  on the orbital period is less than 1\%. However, we do find three very bright outbursts in our ROI. We traced these to dramatic solar flare activity in the intervals MJD~55628--55629, MJD~55991--55996 and MJD~56712--56713. We excised these time intervals from the rest of the analysis. We also inspect probability-weighted light curves of three nearby sources including the blazar PKS~2320$-$035 by folding events within $R=3^\circ$ centered at the source positions. We then perform $H$ tests on the data of the comparison sources using the same selection criteria as used for J2339 and find that none shows significant modulation ($H\le3$). Moreover, the shapes of the light curves of the comparison sources differ from that of J2339. Hence, we conclude that the low-energy modulation is intrinsic to J2339. However it is confined to the on-pulse interval and peaks at pulsar superior conjunction, precisely opposite to expectations for IBS emission beamed along the contact discontinuity.

\subsection{Spectral Analysis} \label{sec:sec2_3}

\begin{figure*}
\centering
\includegraphics[width=5.2 in]{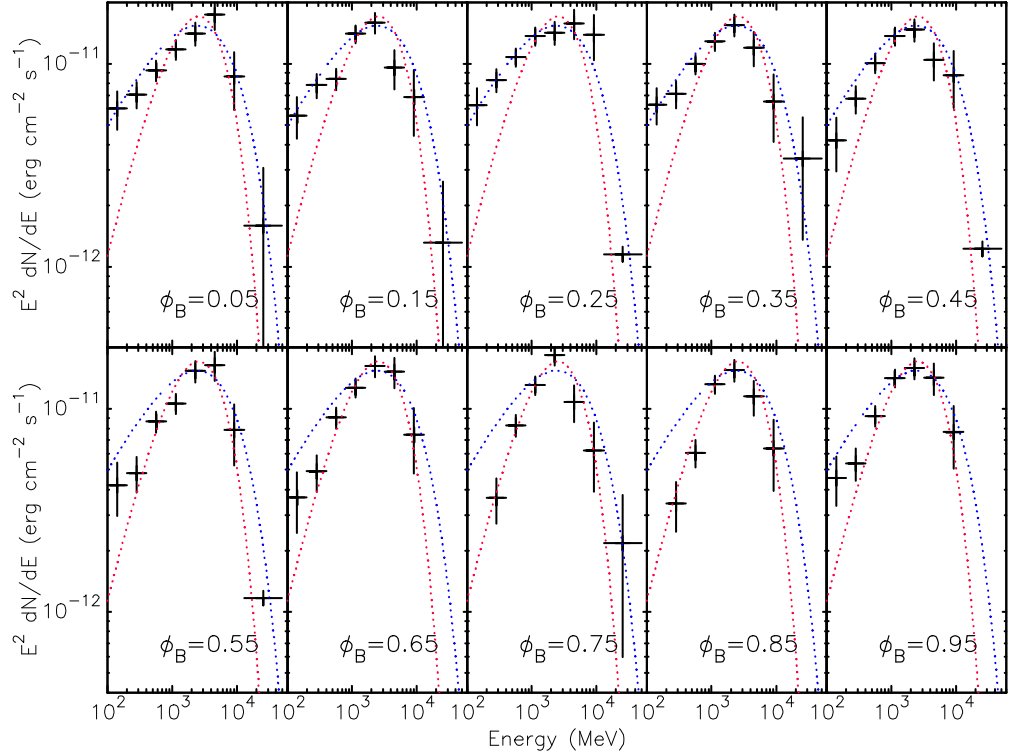}
\figcaption{Phase-resolved SEDs in 10 orbital phase bins. The blue and red dotted lines are the best-fit PLEXP2 models of the orbital-maximum ($\phi_{\rm B}=0.2-0.3$) and the orbital-minimum ($\phi_{\rm B}=0.7-0.8$) spectra, respectively; they are shown for reference.
\label{fig:fig3}
}
\vspace{0mm}
\end{figure*}

\begin{figure*}
\centering
\includegraphics[width=5 in]{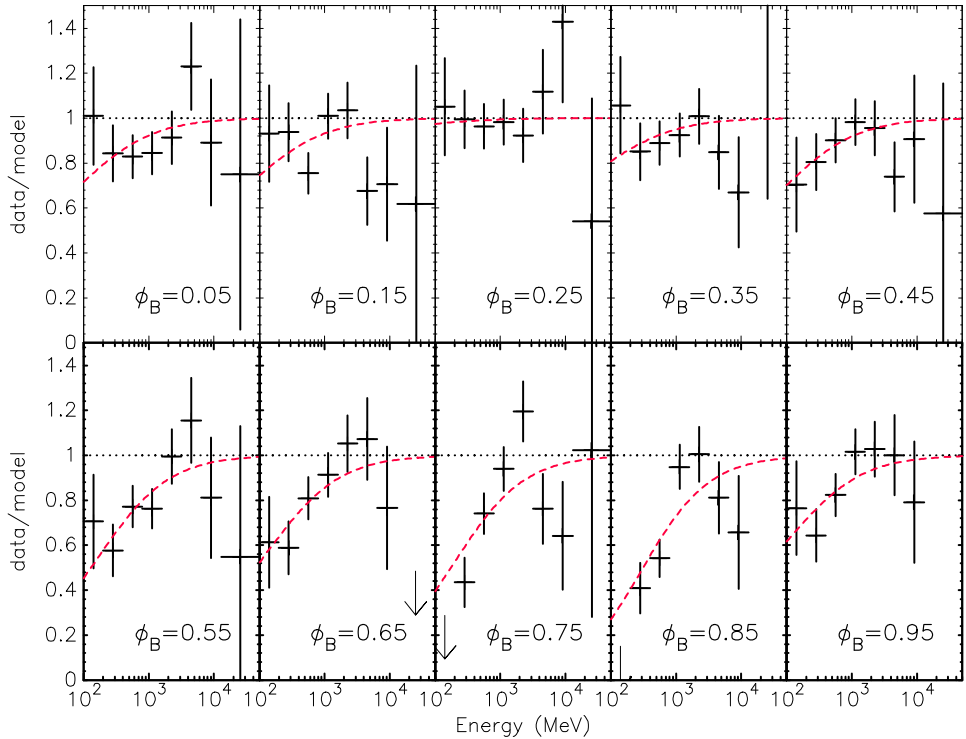}
\figcaption{Phase-resolved SEDs divided by the best-fit orbital-maximum spectral model, and absorption model fits (red dashed; see texts). Horizontal lines (i.e., no absorption) are plotted for reference.
\label{fig:fig4}
}
\vspace{0mm}
\end{figure*}

We next investigate the spectral properties of the source and the modulated flux. We perform binned likelihood analyses in the 100\,MeV--300\,GeV band to measure spectral properties of J2339. We follow the standard binned analysis procedure using the 4FGL model with energy dispersion. Because the ecliptic passes through the RoI, we also include Sun and Moon emission\footnote{https://fermi.gsfc.nasa.gov/ssc/data/analysis/scitools/solar\_t\\emplate.html} and exclude intense solar flare periods (see above) in this analysis. In likelihood fits, we allowed the spectral parameters of J2339, those of brightest field source (PKS~2320$-$035), and the normalizations of the diffuse models ({\tt gll\_iem\_v07} and {\tt iso\_P8R3\_SOURCE\_V2\_v1}) and of the Sun and Moon emission to vary, while other source parameters were  held fixed at the 4FGL model values. We then gradually free parameters of the next brightest sources in the field and compare the fit statistics using the Akaike Information Criteria \citep[AIC;][]{aic74} until improvement of the fit is insignificant.  In all, we find four point sources, the Sun and Moon emission, and the diffuse models require re-optimization.

J2339 is modeled with a power-law exponential cutoff (PLEXP2) model $dN/dE=N_0(E/E_0)^{-\Gamma_{\rm 1}}e^{-aE^{\Gamma_{\rm 2}}}$. In the model, $E_0$ is the reference energy for $N_0$, and $\Gamma_2$ is not well constrained due to the lack of high-energy data for J2339 but $\Gamma_2=0.67$ provides good fits for bright pulsars \citep[][]{fermi4fgl}. We therefore hold these parameters fixed at the 4FGL values $E_0=$1.1\,GeV and $\Gamma_2=$0.67, and note that fixing these parameters does not have a large impact on our investigations below. The best-fit parameters are $N_0=9.5\pm0.8\times 10^{-12}\rm \ ph\ MeV^{-1}\ cm^{-2}\ s^{-1}$, $\Gamma_{\rm 1}=1.06\pm0.07$, and $a=7.8\pm0.6\times 10^{-3}$ with $F_{\rm 0.1-300 GeV}=2.2\pm0.1\times 10^{-8}\rm \ ph\ cm^{-2}\ s^{-1}$; these are fully consistent with the 4FGL values. These values may be affected by small uncertainties due to systematic errors in the effective area and the Galactic diffuse model\footnote{https://fermi.gsfc.nasa.gov/ssc/data/analysis/LAT\_caveats.html}, but their absolute values are not important for our analysis. We note that neither inclusion nor exclusion of the Sun/Moon emission has any significant impact on the results.
	
Because the orbital modulation is pulse-phase dependent, we examine variations in the pulsar spectrum at different orbital phases.  We perform likelihood analyses with on-pulse data in 10 orbital-phase bins, fitting PLEXP2 models and producing spectral energy distributions (SEDs). In the fits, we let the J2339 parameters vary and hold all the other parameters (i.e., for the diffuse and solar/lunar emission, and sources in the RoI) fixed at the phase-averaged values. These are plotted in Figure~\ref{fig:fig3} along with the best-fit spectral models for the orbital-maximum ($\phi_{\rm B}=0.25$, black dotted line) and -minimum bin ($\phi_{\rm B}=0.75$, red dotted line). At orbital maximum the low-energy index $\Gamma_1 (\phi_{\rm B}=0.25)=1.39\pm 0.14$, a fairly typical millisecond pulsar (MSP) LAT spectrum. However the deficit of low-energy counts near orbital minimum means that the PL component is fainter and harder, $\Gamma_1(\phi_{\rm B}=0.75)=0.6\pm0.3$. Note that the $>$GeV flux is nearly constant while, as expected, the low-energy flux varies as described in Section~\ref{sec:sec2_2}.

We investigate if the orbital modulation of the on-pulse spectrum is produced by additional power-law (PL) emission (e.g., from IBS) to the constant pulsar emission which is best represented by the orbital minimum (PLEXP2) spectrum. Because the additional PL parameters cannot be constrained in all the orbital phase bins separately, we apply this composite model first to the orbital-maximum spectrum (holding the PLEXP2 parameters fixed at the orbital-minimum values) in order to determine a representative photon index: $\Gamma=2.6\pm 0.3$. We note that the PL component is detected with test statistic ($TS$) of 25; the freely fit PLEXP2 model has a slightly lower $-\mathrm{log}\mathcal{L}$ ($\Delta\mathrm{log}\mathcal{L}=2$) as compared with the composite model (minimum-fixed PLEXP2 + fit PL), and the latter is not significantly better than the single component model (AIC $p=0.6$).
We then carry out an orbital phase-resolved likelihood analysis with the PLEXP2+PL model by holding the PLEXP2 parameters fixed at those of the orbital minimum and the PL photon index at 2.6. Although the composite model is not significantly better than PLEXP2 in any orbital phase bin, it provides useful parameters for the exploration of physically-motivated models (see Section~\ref{sec:sec3}).

We also measure the orbital-phase-integrated off-pulse spectrum using a PLEXP2 or a PL model in the likelihood analysis. Because the emission is very faint, we fit the normalization of the PLEXP2 and the normalization and index of the PL model, holding all the other parameters fixed at the phase-average values. The emission is detected with $TS$ of 18 and 19 for the PLEXP2 and PL (with an index of $\Gamma=2.4\pm0.2$) models, respectively.

\section{Toy Models for the Orbital Variability} \label{sec:sec3}

We next seek a plausible origin for this modulation. Since the orbital phase evolution is opposite to that expected from an IBS component (and to that of the observed double-peaked X-ray emission), it is highly unlikely that this is standard IBS shock emission. This is underlined by the fact that the modulation is of pulsed photons.  Indeed, it seems to be directly associated with soft counts that follow the pulse profile, as the two spin-phase peaks account for most of the orbital modulation; $P_1$ ($\phi$=0.3--0.33) yields an orbital modulation $H_1$=22 and $P_2$ ($\phi$=0.7--0.75) yields $H_2$=17, compared to an on-pulse $H=40$.

\begin{figure*}
    \centering
    \begin{tabular}{cc}
    \includegraphics[width=3in]{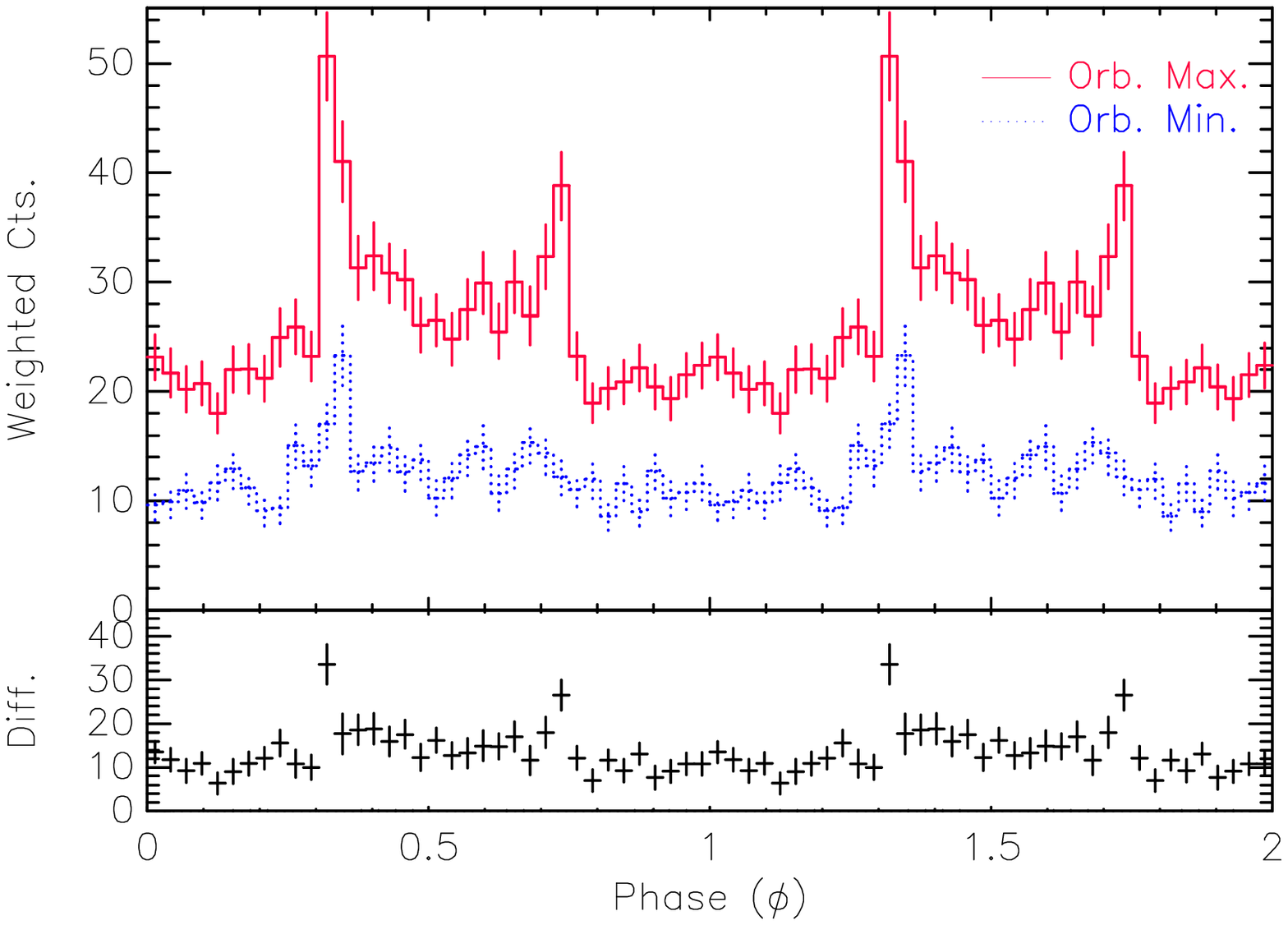} &
    \includegraphics[width=3in]{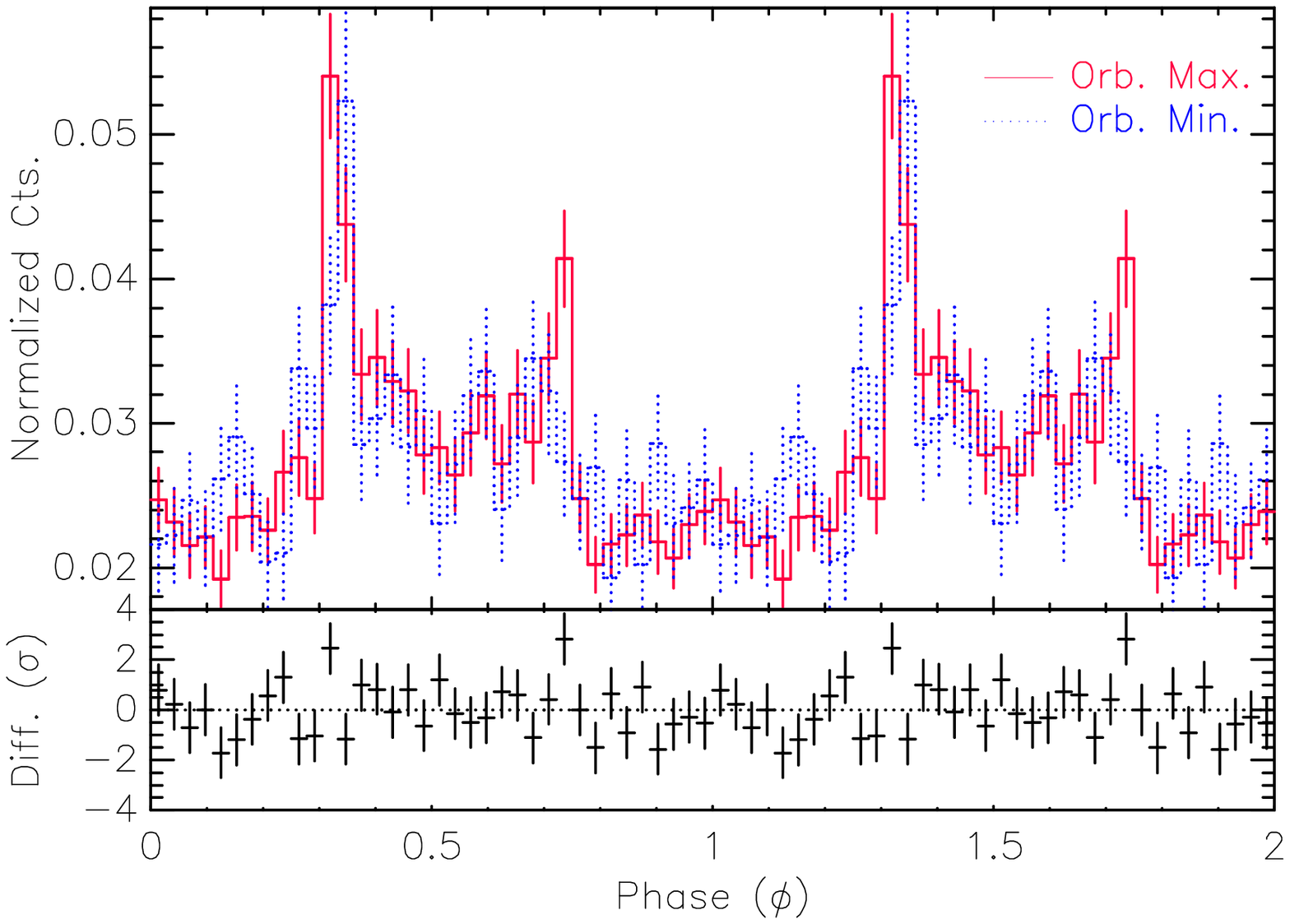}
    \end{tabular}
    \caption{Low-energy (100-600\,MeV) gamma-ray pulse profiles at orbital minimum and maximum (left) and those normalized to the total weighted counts (right). Note that the excess energy at maximum is in both the pulse peaks and bridge, and that at maximum the peak separation appears somewhat larger.  The weights have been corrected for the changing source spectrum as described in the main text. The lower panels display the difference between the profiles in weighted-counts (left) and $\sigma$ (right) units.}
    \label{fig:LowEPulse}Discovery of Orbital
\end{figure*}

 Figure~\ref{fig:LowEPulse} compares the pulse profiles at orbital minimum ($\phi_{\rm B}=0.6-0.9$) and maximum ($\phi_{\rm B}=0.1-0.4$) for the low-energy gamma rays. The stronger pulse at orbital maximum is clearly seen, as well as a small, potential shift in the position and amplitudes of the pulse peaks.

In order to quantitatively compare the pulse profiles at orbital minimum/maximum, we employed the simplest model sufficient to describe the pulse structure, namely a wrapped lorentzian to describe the first peak (P1) and wrapped gaussians for the second peak (P2) and the bridge emission between the two, and we used the log likelihood for this model (Equation \ref{eq:loglike}) to assess the significance of possible shape changes in the pulse.  The weights appearing in the log likelihood are typically calculated for a single spectral model, but here we know the spectral shape and intensity varies between orbital maximum and minimum.  Because the source is brighter, particularly at low energies, during orbital maximum, weights are overestimated at orbital minimum, potentially inducing an unpulsed component.  To adjust the weights to the correct spectrum, we ``re-weight'' each photon $w\rightarrow \alpha w / (\alpha w +1-w)$ with $\alpha$ the flux ratio e.g. $\alpha_{min} = f_{min} / \delta\phi_{min}(f_{min} + f_{max})$ for photons \citep[see][for more details]{kerr2019} from orbital minimum and mutatis mutandis for photons from orbital maximum.  The spectral models obtained in 10 orbital phase bins (\S\ref{sec:sec2_3}) provide sufficient granularity and are commensurate with the $\delta\phi_{\rm B}=0.3$ windows adopted here, so we employ them for the re-weighting.

With the corrected weights, we maximize the likelihood for photons from the orbital maximum to obtain a baseline pulse template.  We then re-fit a subset of parameters, e.g. the location and/or position of one or both of the two peaks, to measure the change in log likelihood and thus measure the significance, which we calculate via Wilks' Theorem, viz. that twice the change in log likelihood under the null hypothesis is distributed as $\chi^2$ in the number of free parameters.  Our results are shown in Table~\ref{ta:pulse}.  We find modest evidence for differences in peak parameters, both individually and collectively.  The most significant is a shift in the positions of P1 and P2 later and earlier in phase, respectively, at about the 3.3\,$\sigma$ level.

To compensate for any discrepancy in the re-normalization of the weights, which might mimic an unpulsed component, we repeat our fits with an additional unpulsed component and measure shape changes relative to it.  We find little evidence for such a component (1.2\,$\sigma$), and the results in Table \ref{ta:pulse} are stable (changes of 0.1--0.2$\sigma$).

\begin{table}
\begin{center}
\caption{Analysis of Variations in Pulse Profile}
\label{ta:pulse}
\begin{tabular}{ l c c }
\hline
\hline
Free Parameter(s) & $ \delta \log \mathcal{L} $ & Significance ($ \sigma$)\\
\hline
P1 Amplitude\dotfill & 0.2 &0.6 \\
P1 Position\dotfill & 3.1 & 2.5 \\
P1 Pos.$+$Amp.\dotfill &  3.3 & 2.1 \\
P2 Amplitude\dotfill & 0.9 & 1.3 \\
P2 Position\dotfill & 2.9 & 2.4 \\
P2 Pos.$+$Amp.\dotfill  & 3.1 & 2.0 \\
P1,P2 Position\dotfill & 6.8 & 3.3 \\
P1,P2 Pos.$+$Amp.\dotfill & 6.1 & 2.4 \\
\hline
\end{tabular}\\
\end{center}
\end{table}

This is remarkable since the standard assumption is that spin-powered pulsar emission is independent of the binary phase. Here with a circular orbit and the companion tidally locked there should be no orbital modulation of conditions in the magnetosphere.  Averaged over a spin period, the pulsar emission should be constant.

Here we explore possible reasons for the orbital pulse flux variation. The first possibility is that the pulsar emission is intrinsically constant, but absorbed at low energy, with an absorption linked to the wind flows. This is attractive since the binary flux maximum ($\phi_{\rm B}=0.25$) spectrum is typical of a LAT MSP. Moreover, the high-energy flux ($\ge2$\,GeV) does not show strong orbital variation. Since orbital minimum at $\phi_{\rm B}\ge 0.75$ is at pulsar inferior conjunction, when we are looking through the pulsar wind-filled channel, the absorption should be associated with the swept-back pulsar wind. This absorption needs to be strongest at low energy. Compton scattering with the Klein-Nishina effect \citep[][]{kl1929} provides such a cross section:

\begin{multline}
\sigma_{\rm KN} = 
\frac{3 \sigma_T}{4} \left[ \frac{1+x}{x^3} \left( \frac{2x(1+x)}{1+2x} -{\rm ln} (1+2x) \right) \right. \\
\left. + \frac{1}{2x}{\rm ln} (1+2x) - \frac{(1+3x)}{(1+2x)^2} \right],
\end{multline}
where $\sigma_T$ is the Thomson scattering cross section and $x=E_\gamma/m_ec^2$. Remembering that during the minimum phases we are looking along the flow of the post-termination shock (but pre-contact discontinuity) pulsar wind toward the observer, with bulk motion $\Gamma_{\rm w}$, we may take  $x=E_\gamma/(\Gamma_{\rm w}m_ec^2)$. The scattering cross section decreases as $\sim 1/E_\gamma$ (until pair production of the electron field dominates at higher energies, when it is energy independent.)

\begin{figure*}[t]
\centering
\begin{tabular}{cc}
\includegraphics[width=3 in]{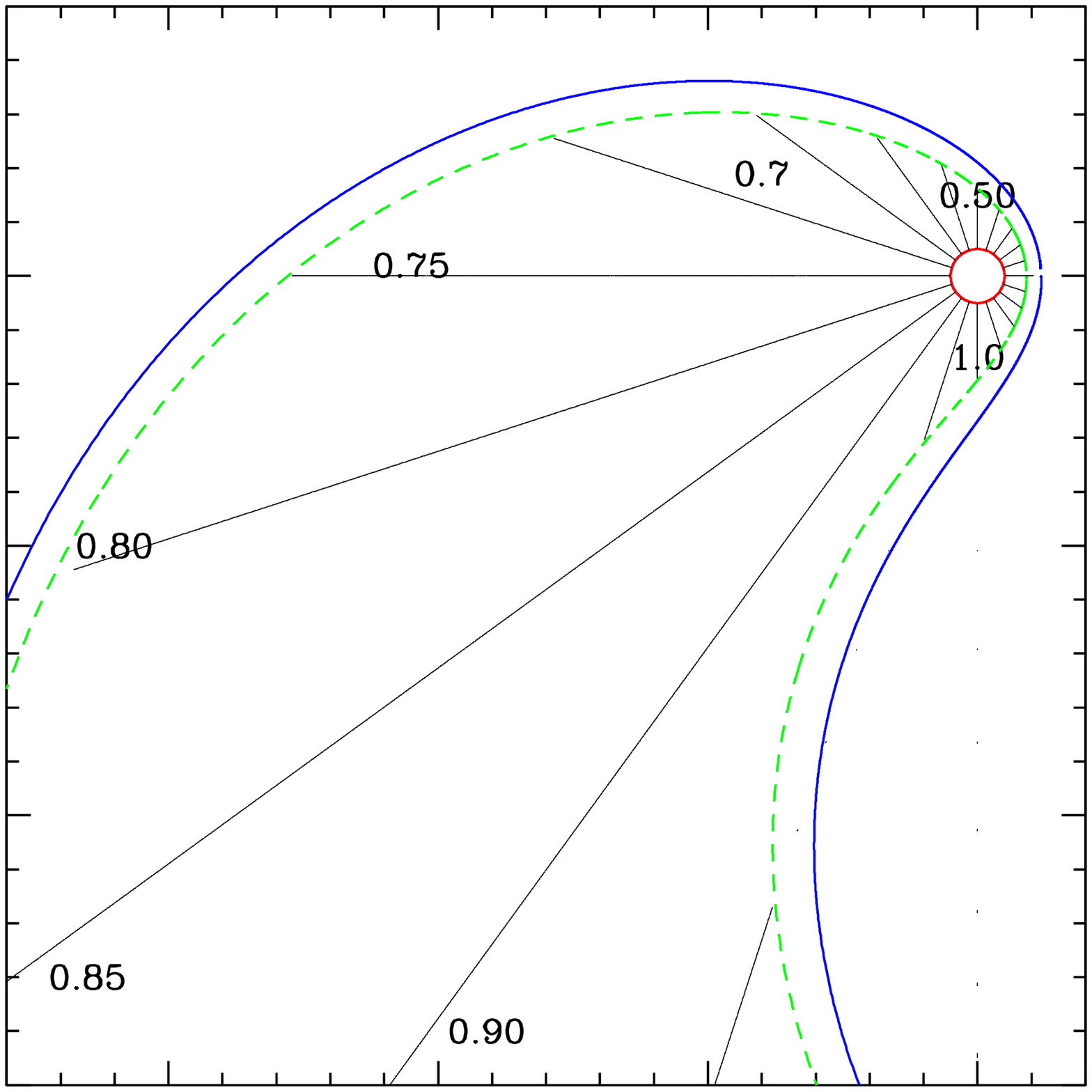} &
\hspace{0.5 cm}
\includegraphics[width=3 in]{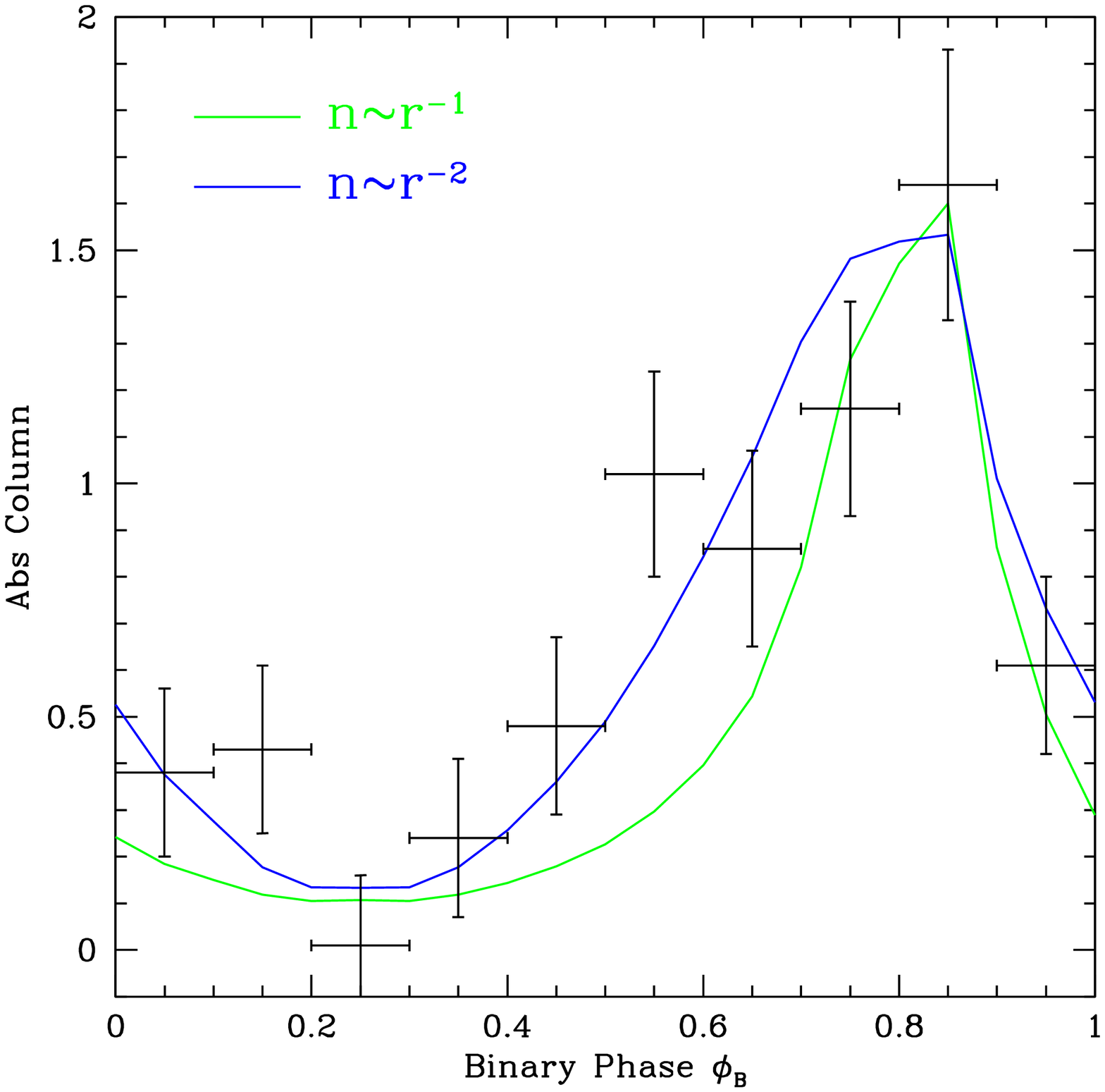} \\
\end{tabular}
\figcaption{A toy model for Compton scattering by IBS electrons.
Left: the system geometry for a swept back IBS, with the contact discontinuity (dashed green), the pulsar wind shock (solid red) and the stellar wind shock (solid blue) schematically shown. The Earth lines-of-sight to the pulsar (black lines) are shown with labeled $\phi_{\rm B}$ for several phases. Note that the sweepback ensures that the largest column inside the shocked pulsar wind occurs for $\phi_{\rm B}>0.75$. 
Right: the column density for the 3-D ($r^{-2}$) and 2-D ($r^{-1}$) cases, starting from the pulsar wind shock stand-off radius. This is compared to the absorption column fit (Fig.~\ref{fig:fig4}) to the orbital modulation data (arbitrary normalization for the y axis).
\label{fig:fig5}
}
\vspace{-0mm}
\end{figure*}

To implement such a model, we assume a bulk Lorentz factor $\Gamma_{\rm w}=100$ and model the orbital phase-resolved SEDs as a (fixed) pulsar PLEXP2 function absorbed by $e^{\tau(E)}$ with $\tau(E)= C(\phi_{\rm B})\sigma_{\rm KN}(E)$ and the effective absorption column $C(\phi_{\rm B})$ varying with orbital phase. All intrinsic pulsar spectral parameters are held fixed at orbital-maximum fit values, leaving the effective absorption column $C(\phi_{\rm B})$ as the only adjustable parameter. Instead of performing likelihood analyses for each phase bin, we divide the measured SEDs (Fig.~\ref{fig:fig3}) by the fixed (orbital maximum) pulsar model (Fig.~\ref{fig:fig4}), and vary $C(\phi_{\rm B})$ to fit the ratio in each phase bin. The values are plotted in Figure~\ref{fig:fig5} right panel and compared to a simple geometrical estimate of the variation in the column density through the $e^+/e^-$ of the post-shock pulsar wind. We can assume that the wind expands ($n \sim 1/r^2$) or is approximately equatorial ($n \sim 1/r$). The former gives a somewhat wider peak and better match to the data. The inferred column density $C(\phi_{\rm B})$ is maximal at $\phi_{\rm B}\approx 0.85$, in reasonable agreement with the pattern from IBS sweepback. Note that we assume that the scattering column starts at a termination shock distance comparable to the standoff of the pulsar wind at the nose; if the denser pair plasma closer to the pulsar dominated, this constant absorption would wash out any orbital phase variation, so the dominant absorption must occur on the scale of the orbital separation $a\sim 2R_\odot$.

However, for this model to make any sense we must reach a maximum optical depth $\tau\sim 1$ at an observed $E_\gamma \approx 0.1$\,GeV. Suppose that $10^{34} {\dot E}_{34}\ {\rm erg\, s^{-1}}$ of the pulsar spindown luminosity \citep[total power ${\dot E}_{34}=2.1$ considering the Shklovskii effect for J2339;][]{gaia18,jkcc+18} is converted to pairs in a cold wind with bulk Lorentz factor $\Gamma_{\rm w}$. This wind is axisymmetric around the pulsar until the closest termination shock at the nose $r_0 \approx a/3$ where $a =10^{11}a_{11}$\,cm is the orbital separation, so for the pairs to produce an orbital modulation, the bulk of the absorption must be created at or beyond this distance, e.g. by conversion of a high $\sigma$ (magnetization) wind. For these assumptions, the Thomson scattering depth from a radius $a$ through this pair plasma is $\tau \approx \sigma_T {\dot E}/(4\pi a m_e c^3 \Gamma_{\rm w}) \approx 7 \times 10^{-7} {\dot E}_{34}/(a_{11}\Gamma_{\rm w})$. Since the Klein-Nishina cross section is reduced by $\sim E_\gamma/(\Gamma_{\rm w} m_e c^2)$ at gamma-ray energies, this misses by several orders of magnitude. Geometry might increase the column by a factor of a few if, e.g. the Earth line-of-sight happens to pass through a $\sim$2-D electron layer. One might also imagine that neutral H penetrates the IBS and is ionized by the pre-shock pulsar wind. However, even using all spindown power for ionization rather than pair production only increases the $e^-$ density by $1.02\times 10^6 {\rm eV}/13.6{\rm eV} \approx 8 \times 10^4$. Thus it seems impossible for the pulsar to produce sufficient optical depth distributed over the orbital length scale, as required for an orbitally modulated scattering or absorption. 

\begin{figure}
    \centering
    \includegraphics[width=3in]{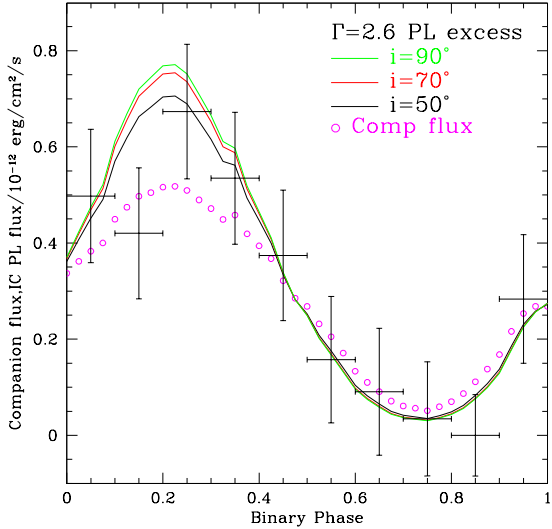}
    \caption{Magenta points show the estimated companion's optical flux (arbitrary normalization) as a function of phase \citep[taken from][]{kra19} while the lines estimate the associated ICS power (arbitrary normalization) produced in the near-pulsar wind, for three different inclinations. The black points (left scale) are from a simple $\Gamma=2.6$ PL fit to the flux excess above the orbital minimum ($\phi_{\rm B}=0.75$) PLEXP2 pulsar spectrum (Section~\ref{sec:sec2_3}).}
    \label{fig:ICSemis}
\end{figure}

The alternative is to add {\it pulsed} flux to the orbital minimum. Here the likely mechanism involves Compton up-scatter of low-energy photons from the heated companion \citep[e.g.,][]{wtch+12,arjk+17,ntsl+18}. Such up-scattered photons are believed to dominate the orbitally modulated signal from high-mass gamma-ray binaries \citep[][]{d13}. The Compton power per electron upscattered from a soft photon energy density $u_*$ by the pulsar wind of bulk motion $\Gamma_{\rm w}$ is
$$ 
P_{\rm ICS} = \sigma_T c \, u_* (1-\beta_{\rm w}\mu)[(1-\beta_{\rm w}\mu)\Gamma_{\rm w}^2 -1]
$$
with $\mu=\cos\theta_{\rm ICS}$ the angle between the pulsar wind $e^+/e^-$ (radial toward Earth in the pre-shock flow) and the stellar photons. Here we have the advantage that the power modulation is imparted by the asymmetric (day-night) radiation from the companion, so that the gamma rays may be produced close to the pulsar. Figure~\ref{fig:ICSemis} illustrates an ICS excess model, with the points showing the companion flux as a function of phase, and a simplified estimate of the associated ICS signal, compared with a PL fit to the orbitally varying LAT component.  

We can use either the observed magnitudes $m_r=18.8$ (and the fit $d\approx 1.25$\,kpc source distance) or the fit stellar temperature $T_{\rm eff}=4800$\,K to estimate the photon energy density from the companion \citep[][]{rs11,kra19}. In the near zone of the pulsar wind (about 2$R_\odot$ from the companion) this is $u_*\approx 0.2\ {\rm erg\, cm^{-3}}$. As above the pulsar produces electrons at a rate ${\dot N_e} \approx {\dot E}/(\Gamma_{\rm w} m_e c^2) \approx 2 \times 10^{40}/\Gamma_{\rm w} \, e^-/{\rm s}$. Assuming that the Compton upscattering occurs over a distance $\delta r_\odot$ (in Solar radius units) then the ICS luminosity at orbital maximum is
$$L_{ICS,max} \approx \sigma_T c u_\ast 4 \Gamma_{\rm w}^2 {\dot N_e} (\delta r/c) \approx 9 \times 10^{26} \Gamma_{\rm w} \delta r_\odot {\rm erg\ s^{-1}}.$$
This is to be compared with the observed power-law flux, which contributes $\sim 5 \times 10^{-13}\ {\rm erg\ cm^{-2}\ s^{-1}}$ of flux at maximum in a $dN_\gamma/dE_\gamma = K E_\gamma^{-2.6}$ component (Figure~\ref{fig:ICSemis}) of which we found a hint in our data fits ($TS$=25) although an addition of this power-law component is not strongly favored over a simple freely fit PLEXP2 model, providing a modest AIC value of $p=0.6$ (Section~\ref{sec:sec2_3}). At the fit source distance, this is $\sim 2 \times 10^{31}\ {\rm erg\ s^{-1}}$, assuming beaming into $\pi$ steradians. Comparing with $L_{ICS,max}$, we see that such flux can be produced if $\Gamma_{\rm w} \delta r_\odot = 2 \times 10^4$.

Thus this model appears energetically feasible. However there are a number of less than satisfactory aspects. First, the underlying PLEXP2 model is unusually hard for a LAT pulsar. In fact at $\phi_{\rm B}=0.75$ we fit $\Gamma_1 = 0.6\pm0.3$. This is comparable to the smallest values seen for LAT pulsars. Thus in this model we are adding a very soft $\Gamma=2.6$ variable ICS component (power law) to an unusually hard MSP spectrum (Section~\ref{sec:sec2_3}). This is in contrast to the absorption picture, where the PLEXP2 spectrum at maximum is quite typical of LAT MSP. Also we require a large fraction of the spin-down power to go into $e^+/e^-$ in the pre-shock wind. This means that it must become particle dominated (low $\sigma$) relatively early. Since $\Gamma_{\rm w} \sim 10^4$ is inferred for the bulk PWN flow in other pulsars \citep[e.g.,][]{kabr12}, we also infer that it stays in this state for $\delta r_\odot \approx 1$, i.e. a good fraction of the way to the IBS termination shock. The ICS estimate above assumes a spherical wind, so an equatorial concentrated wind can help modestly decrease the power requirements. Note that this only helps if the Earth line-of-sight is covered by this equatorial wind flow -- with the optically fit $i\approx 69^\circ$ \citep[][]{kra19} this concentration is not a large help. It does suggest that interacting binary pulsars observed at large $i$ would be more likely to show ICS orbital modulation. 

Most importantly, to keep this emission pulsed in phase with the classic magnetospheric pulsations, the scattering electrons need to be in a precisely phased structure, presumably the striped pulsar wind, over this large distance.  

\section{Discussion and Conclusions}
\label{sec:sec4}

We have discovered low-energy gamma-ray orbital modulation of J2339 emission in 11\,yr of {\it Fermi}-LAT data. The modulation is strongest in the ``on-pulse'' interval. The significance of modulation increases approximately monotonically with time, and variability of exposure and/or nearby bright sources does not explain the modulation. We therefore conclude that the modulation is intrinsic to J2339, and phased with the pulsed magnetospheric emission.

The origin of this pulsed flux is not clear. Klein-Nishina limited scattering of the magnetospheric emission by electrons in the pulsar wind seems attractive in that it naturally explains the low-energy dominance and the orbital phase variation of the modulation by linking the scattering depth to the column of electrons in the trailed pulsar wind. And of course a simple energy-dependent attenuation of the pulsed emission is then expected. But the optical depths provided by the expected pulsar pair emission are too low by orders of magnitude. Identifying the modulation as Compton upscattering of companion day-face photons seems energetically more promising although orbitally-varying Compton emissivity due to changes in the scattering angle needs to be taken into account for more detailed modeling. But here the challenge is to ensure that the upscattered photons stay in phase with the pulsed emission. This may be possible in a striped wind extending nearly unperturbed to the IBS termination shock, but it is unclear why the ICS ``pulse profile'' should be so similar to the standard pulsations originating near the pulsar. Additionally, the standard pulsed emission in this scenario is unusually hard, with a very soft component added by the ICS emission. We conclude that additional observations and modeling will be needed to tease out the origin of this remarkable pulse modulation.

Although it is bright, allowing detailed $\gamma$/X/optical study, and relatively strongly heated, J2339's properties do not seem exceptionally unusual. Thus we might expect to find similar pulse modulation in other spider binaries. Indeed, orbital modulation of gamma-ray flux in pulsar binaries has been reported in a few systems \citep[PSR~J1311$-$3430, PSR~J2241$-$5236, and 3FGL~J2039.6$-$5618;][]{arjk+17,ark18,ntsl+18}. This is generally attributed to IBS emission, with synchrotron emission or inverse-Compton upscattering of stellar photons in the IBS as the suggested production mechanism. In both cases we expect the gamma-ray emission to be beamed along with the bulk motion of the IBS electrons, thus similar to the IBS X-rays (generally away from the MSP for BW, away from the companion for RB). However, if Compton upscattering is strongest at the IBS apex before the shocked pulsar wind has joined the bulk flow, it may be directed away from the MSP in either case. Of course for these IBS models we do not expect the excess gamma-ray emission to be pulsed. This would make IBS gamma-ray detection easiest in the off-pulse phases. 

In practice gamma-ray modulation is not always best seen off-pulse. So it seems likely that a pulsed component, like that discovered here, is present in other objects. For example 3FGL~J2039.6$-$5618 shares some properties with J2339; both of them are RBs, and its recently-detected gamma-ray maximum occurs near the optical minimum, opposite to the expected IBS phase. It will be interesting to see if 3FGL~J2039.6$-$5618 also exhibits orbital modulation in the ``on-pulse'' interval. More examples, and more detailed modeling, will certainly help us trace the origin of this phenomenon. Continuously collecting gamma-ray data with the {\it Fermi} LAT and the future {\it AMEGO} telescope \citep[][]{amego19} may give us new insights into pulsar binaries.

\bigskip
\bigskip
\bigskip
\acknowledgments
The \textit{Fermi} LAT Collaboration acknowledges generous ongoing support
from a number of agencies and institutes that have supported both the
development and the operation of the LAT as well as scientific data analysis.
These include the National Aeronautics and Space Administration and the
Department of Energy in the United States, the Commissariat \`a l'Energie Atomique
and the Centre National de la Recherche Scientifique / Institut National de Physique
Nucl\'eaire et de Physique des Particules in France, the Agenzia Spaziale Italiana
and the Istituto Nazionale di Fisica Nucleare in Italy, the Ministry of Education,
Culture, Sports, Science and Technology (MEXT), High Energy Accelerator Research
Organization (KEK) and Japan Aerospace Exploration Agency (JAXA) in Japan, and
the K.~A.~Wallenberg Foundation, the Swedish Research Council and the
Swedish National Space Board in Sweden.
 
Additional support for science analysis during the operations phase is gratefully
acknowledged from the Istituto Nazionale di Astrofisica in Italy and the Centre
National d'\'Etudes Spatiales in France. This work performed in part under DOE
Contract DE-AC02-76SF00515.

This research was supported by Basic Science Research Program through the National Research Foundation of Korea (NRF) funded by the Ministry of Science, ICT \& Future Planning (NRF-2017R1C1B2004566). R.W.R. was supported in part by NASA grant 80NSSC17K0024. Work at NRL is supported by NASA, in part by Fermi Guest Investigator grant NNG19OB19A.

\vspace{5mm}
\facilities{Fermi-LAT}
\software{Fermi ST, Tempo2, PINT}

%\vspace{5mm}
%\facilities{HST(STIS), Swift(XRT and UVOT), AAVSO, CTIO:1.3m, CTIO:1.5m,CXO}
%\software{astropy \citep{2013A&A...558A..33A},  
%          Cloudy \citep{2013RMxAA..49..137F}, 
%          SExtractor \citep{1996A&AS..117..393B}
%          }
\bigskip
\bigskip
\bibliographystyle{apj}
%\begin{thebibliography}{}
\expandafter\ifx\csname natexlab\endcsname\relax\def\natexlab#1{#1}\fi

%\end{thebibliography}

\end{document}